\PassOptionsToPackage{unicode=true}{hyperref} 
\PassOptionsToPackage{hyphens}{url}
\documentclass{llncs}

\usepackage{hyperref}
\hypersetup{
            pdfborder={0 0 0},
            breaklinks=true}

\title{When You Should Use Lists in Haskell\\ (Mostly, You Should
  Not)}
\author{Johannes Waldmann}
\institute{F-IMN, HTWK Leipzig, Germany\\
  \url{https://www.imn.htwk-leipzig.de/~waldmann/}
}

\begin{document}
\maketitle

\begin{abstract}
  We comment on the over-use of lists in functional programming.
  With this respect, we review history of Haskell and some of its libraries,
  and hint at current developments.
\end{abstract}

\section{Introduction}\label{introduction}

It seems that the designers of the programming language
Haskell~\cite{haskellreport}
were madly in love with singly
linked lists. They even granted notational privileges: e.g., the data
constructor for a non-empty list (``cons'') uses just one letter (\texttt{:}).
Looking at the Haskell standard, it also seems that singly
linked lists are more important than the concept of static typing.
When declaring the type of an identifier, we must use two letters
\texttt{::}, because the natural (that is, mathematical) notation
\texttt{:} is taken. 

One goal of language design is to make typical expected usage easy to
write. Once the design is cast in stone, this becomes self-fulfilling:
unsuspecting users of the language will write programs in what they
think is idiomatic style, as suggested by built-in syntactical
convenience and sugar. Haskell lists have plenty of both: convenience by
short notation, and sugar in the form of list comprehensions.
So it seems that it is typical Haskell to not declare types for identifiers,
and use lists all over the place.

I will explain on the following pages
that the purported connection between functional programming and lists
is largely historical, and is detrimental to teaching and programming.
Lists do indeed serve a purpose, but a rather limited one:

\begin{itemize}
\item
  \emph{iteration}: you use each element of a collection \emph{in order}
  (FIFO or LIFO), on demand, and at most once.
\end{itemize}

If you use lists for anything else, in particular

\begin{itemize}

\item
  \emph{storage}: you use elements \emph{out of order}, e.g., by known
  index, or you access (``search'') elements by a property,
\end{itemize}

then you're doing it wrong, \emph{wrong}, \emph{wrong}. I will explain
why, and what you should do instead.

As an appetizer, here are some bumper-sticker slogans extracted from the
sections that follow.

\begin{itemize}
\item
  If your program accesses a list by index (with the \texttt{!!}
  operator), then your program is wrong. (Section~\ref{what-lists-are-not-good-at})
\item
  If your program uses the \texttt{length} function, then your program
  is wrong. 
\item
  If your program sorts a list, then your program is wrong.
\item
  If you wrote this \texttt{sort} function yourself, then it is doubly
  wrong.
\item
  The ideal use of a list is such that will be removed by the compiler.
  (Section~\ref{lists-that-are-not-really-there})
\item
  The enlightened programmer writes list-free code with
  \texttt{Foldable}. (Section~\ref{lists-that-are-really-not-there})
\end{itemize}

I will not enter arguments about syntax. (The notation of \emph{cons}
as ``\verb|:|'', and type declarations with ``\verb|::|'',
came to Haskell from Miranda~\cite{DBLP:journals/eatcs/Turner87},
and has recently been reversed in Agda~\cite{norell:thesis}.)
This paper is about the semantics (and consequently, performance)
of singly linked lists, as we have them in Haskell.

Note that the name \verb|List| has a different meaning in, e.g.,
the  Java standard library~\cite{java.util.List},
where it denotes the abstract data type of sequences
(collections whose elements can be accessed via a numerical index).
They have linked lists as one possible realization,
but there are others, for which \verb|length| and \verb|!!| (indexing)
are not atrocious. Of course, there are Haskell libraries
for efficient sequences~\cite{hackage:vector}.

\paragraph{Acknowledgments}\label{acknowledgments}

This text was originally published (in 2017) at
\url{https://www.imn.htwk-leipzig.de/~waldmann/etc/untutorial/list-or-not-list/}

Joachim Breitner, Bertram Felgenhauer, Henning Thielemann and Serge Le
Huitouze have sent helpful remarks on earlier versions of
this text, and I thank WFLP reviewers for detailed comments.
Of course, all remaining technical errors are my own, and I
do not claim the commenters endorse my views.

\section{Where Do All These Lists Come From?}\label{where-do-all-these-lists-come-from}

Certainly Haskell builds on the legacy of
LISP~\cite{DBLP:journals/cacm/McCarthy60,graham:roots-of-lisp}, the very first
functional programming language - that has lists in its name already.
Indeed, singly linked lists were the only way to structure nested data
in LISP.

This is a huge advantage (it's soo uniform), but also a huge
dis-advantage (you cannot distinguish data - they all look like nested
lists). For discriminating between different shapes of nested data,
people then invented algebraic data types, pattern matching, and static
typing. The flip side of static typing is that it seems to hinder
uniform data processing. So, language designers invented
polymorphism~\cite{hindley:principal} and generic programming,
to get back flexibility and
uniformity, while keeping static safety.

This was all well known at the time of Haskell's design, so why are
lists that prominent in the language? The one major design goal for
Haskell was: to define a functional programming language with static
typing and \emph{lazy evaluation}. For functional programming with
static typing and strict evaluation, the language ML was already well
established~\cite{DBLP:conf/sfp/Turner12}. Haskell has
lazy evaluation for function calls, including constructor applications,
and a prime showcase for this is that we can handle apparently infinite
data, like streams:

\begin{verbatim}
data Stream a = Cons a (Stream a)
\end{verbatim}

Such streams do never end, but we can still do reasonable things with
them, by evaluating finite prefixes on demand, in other words, lazily.

Now a standard Haskell list is just such a stream that \emph{can} end,
because there is an extra constructor for the empty stream

\begin{verbatim}
data Stream a = Nil | Cons a (Stream a)
\end{verbatim}

and the problem with Haskell lists is that people tend to confuse what
the type \texttt{{[}\ a\ {]}} stands for:

\begin{itemize}

\item
  is it: a lazy (possibly infinite) stream (yes)
\item
  or is it: a random-access sequence (no); or a set (hell no)
\end{itemize}

If you have an object-oriented background, then this is the confusion

\begin{itemize}

\item
  between an iterator (that allows to traverse data)
\item
  and its underlying collection (that holds the actual data).
\end{itemize}

A Haskell list is a very good iterator - we even might say, it is the
\emph{essence} of iteration, but it makes for a very bad implementation
of a collection.

\section{And Where Are These Lists Today?}\label{and-where-are-these-lists-today}

The (historic) first impression of ``functional programming is
inherently list processing'' seems hard to erase. Indeed it is
perpetuated by some Haskell teaching texts, some of which are popular.
If a text(book) aimed at beginners presents \texttt{Int},
\texttt{String} and \texttt{{[}Int{]}} on pages 1, 2, 3; but keywords
\texttt{data} and \texttt{case} (and assorted concepts of algebraic data
type and pattern matching) appear only after the first two hundred
pages, well, then something's wrong.

For example, the shiny new (in 2015) \url{http://haskell.org/} gets this exactly
wrong. At the very top of the page, it presents example code that is
supposed to highlight typical features of the language. We see a version
of Eratosthenes' sieve that generates an infinite list of primes.
Declarative it certainly is. But see - Lists, numbers. Must be typical
Haskell.

Also, that example code features list comprehensions - that's more
syntactic sugar to misguide beginners. And it claims ``statically typed
code'' in the previous line, but - there is no type annotation in the
code. Yes, I know, the types are still there, because they are inferred
by the compiler, but what does a beginner think?
The icing on all this is that the one thing that this example code
really exemplifies, namely: infinite streams with lazy evaluation, is
not mentioned anywhere in the vicinity.

\section{What Lists Are (Not) Good At}\label{what-lists-are-not-good-at}

The answer is simple: \emph{never use lists for data storage with
out-of-order access.}

Accessing an element of a singly linked list at some arbitrary position
\(i\) takes time proportional to \(i\), and you do not want that.

The only operations that are cheap with a singly linked list are
\begin{itemize}
\item accessing its first element element (the head of list),
\item and prepending a fresh first element.
\end{itemize}

So can we access any of the other elements in any reasonable way? Yes,
we can access the second element, but only after we drop the first one.
The second element becomes the ``new head'', and the ``old head'' is
then gone for good. We are singly linked, we cannot go back. If we want
to access any element in a list, we have to access, and then throw away,
all the elements to the left of it.

In other words, lists in Haskell realize
the \emph{iterator} design pattern~\cite{gof:design-patterns}
(Iterator in Java, IEnumerator in C\#). So - they are a means of expressing
control flow. As Cale Gibbard (2006) puts it~\cite{gibbard:lists-are-loops},
``lists largely take the
place of what one would use loops for in imperative languages.''

Or, we can think of a list as a stack, and going to the next element is
its ``pop'' operation.

Can we really never go back? Well, if \texttt{xs} is an iterator for
some collection, then \texttt{x\ :\ xs} is an iterator that first will
provide \texttt{x}, and after that, all elements of \texttt{xs}. So,
thinking of a list as a stack, the list constructor (\texttt{:})
realizes the ``push'' operation.

On the other hand, the ``merge'' function (of ``mergesort'' fame, it
merges two monotonically increasing sequences) is a good use case for
Haskell lists, since both inputs are traversed in order (individually -
globally, the two traversals are interleaved). Its source code is

\begin{verbatim}
merge :: Ord a => [a] -> [a] -> [a]
merge as@(a:as') bs@(b:bs')
  | a `compare` b == GT = b:merge as  bs'
  | otherwise       = a:merge as' bs
merge [] bs         = bs
merge as []         = as
\end{verbatim}

We never copy lists, and we only ever pattern match on their beginning.

Indeed you can easily (in Java or C\#) write a \texttt{merge} function
that takes two iterators for monotone sequences, and produces one. Try
it, it's a fun exercise. Use it to learn about \texttt{yield\ return} in
C\#.

Back to Haskell: how should we implement merge-sort then, given
\texttt{merge}? The truthful answer is: we should not, see below. But
for the moment, let's try anyway, as an exercise. A straightforward
approach is to split long input lists at (or near) the middle:

\begin{verbatim}
msort :: Ord a => [a] -> [a]
msort [] = [] ; msort [x] = [x]
msort xs = let (lo,hi) = splitAt (div (length xs) 2) xs
           in  merge (msort lo) (msort hi)
\end{verbatim}

but this is already wrong. Here, \texttt{xs} is not used in a linear
way, but twice: first in \texttt{length\ xs} (to compute to position of
the split), and secondly, in \texttt{splitAt\ ...\ xs}, to actually do
the split. This is inefficient, since the list is traversed twice, which
costs time -- and space, since the spine of the list will be held in
memory fully.

Indeed, the Haskell standard library contains an implementation of
mergesort that does something different. The basic idea is

\begin{verbatim}
msort :: Ord a => [a] -> [a]
msort xs = mergeAll (map (\x -> [x]) xs)
  where mergeAll [x] = x
        mergeAll xs  = mergeAll (mergePairs xs)
        mergePairs (a:b:xs) = merge a b: mergePairs xs
        mergePairs xs       = xs
\end{verbatim}

But the actual implementation~\cite{hackage:list}
has an extra provision for handling monotonic segments
of the input more efficiently. Make sure to read the references
mentioned in the comments atop the source code of \verb|Data.OldList.sort|,
including~\cite{augustsson:heap-sort}.

But - if you sort a list in your application \emph{at all}, then you're
probably doing something wrong already. There's a good chance that your
data should have \emph{never} been stored in a list from the start.

How \emph{do} we store data, then?
This is \emph{the} topic of each and every ``Data Structures and Algorithms'' course,
and there \emph{are} corresponding Haskell libraries.
For an overview, see~\cite{waldmann:data}.

\section{Is There No Use For Lists As Containers?}\label{is-there-no-use-for-lists-as-containers}

There are two exemptions from the rules stated above.

If you are positively certain that your collections have a
\emph{bounded} size throughout the lifetime of your application, then
you can choose to implement it in any way you want (and even use lists),
because all operations will run in constant time anyway. But beware -
``the constant depends on the bound''. So if the bound is three, then
it's probably fine.

Then, \emph{education}. With lists, you can exemplify pattern matching
and recursion (with streams, you exemplify co-recursion). Certainly each
student should see and do recursion early on in their first term.

But the teacher must make very clear that this is in the same ballpark
with exercises of computing the Fibonacci sequence, or implementing
multiplication and exponentiation for Peano numbers: the actual output
of the computation is largely irrelevant. What is relevant are the
techniques they teach (algebraic data types, pattern matching,
recursion, induction) because \emph{these} have broad applications.

So, indeed, when I teach~\cite{waldmann:how-i-teach}
algebraic data types, I do also use singly linked lists,
among other examples (Bool, Maybe, Either, binary trees),
but I always write (and have students write)

\begin{verbatim}
data List a = Nil | Cons a (List a)
\end{verbatim}

I do avoid Haskell's built-in lists as long as possible, that is,
until students understand their use (representing iterators)
in the application programmer interfaces of recommended libraries
like \verb|Data.Set| and \verb|Data.Map|.

\section{Lists That Are Not Really There}\label{lists-that-are-not-really-there}

We learned above that the one legitimate use case for lists is
iteration. This is an important one. It's so important that the
Glasgow Haskell compiler (GHC) applies
clever program transformations~\cite{DBLP:conf/fpca/GillLJ93}
in order to generate efficient machine code for programs that
combine iterators (that is, lists) in typical ways.

The idea is that the most efficient code is always the one that is not
there at all. Indeed, several program transformations in the compiler
remove intermediate iterators (lists). Consider the expression

\begin{verbatim}
sum $ map (\x -> x*x) [1 .. 1000]
\end{verbatim}

which contains two iterations: the list \texttt{{[}1\ ..\ 1000{]}} (the
\emph{producer}), and another list (of squares, built by \texttt{map},
which is a \emph{transformer}). Finally, we have \texttt{sum} as a
\emph{consumer} (its output is not a list, but a number).

We note that a transformer is just a consumer (of the input list)
coupled to a producer (of the output list). Now the important
observation is that we can remove the intermediate list between a
producer and the following consumer (where each of them may actually be
part of a transformer).

This removal is called \emph{fusion}. With fusion, lists are used
prominently in the source code, but the compiler will transform this
into machine code that does not mention lists \emph{at all}! In
particular, the above program will run in constant space, as it allocates
\emph{nothing} on the heap, and it needs \emph{no} garbage collector.

For historic reference and detail, see~\cite{DBLP:journals/tcs/Wadler90}.

\section{Lists That Are Really Not There}\label{lists-that-are-really-not-there}

Now I will present a more recent development (in the style of writing
libraries for Haskell) that aims to remove lists from the actual \emph{source} code.
Recall that deforestation removes lists from \emph{compiled} code.

We have, e.g., \verb|and [True,False,True]=False|, and
you would think that \texttt{and} has type
\verb|[Bool] -> Bool|, as it maps a list of Booleans
to a Boolean. But no, the type is
\verb|and :: Foldable t => t Bool -> Bool|,
and indeed we can apply \texttt{and} to any object of a type
\texttt{t\ a} where \texttt{t} implements \texttt{Foldable}.

The constraint \verb|Foldable t|
for a type constructor \verb|t| of kind \verb|* -> *|
means that there is an associated function
\verb|toList :: t a -> [a]|. So, \texttt{c} of
type \texttt{t\ a} with \texttt{Foldable\ t} can be thought of as some
kind of collection, and we can get the stream of its elements by
\verb|toList c|. This is similar to \verb|c.iterator()| for
\verb|Iterable<A> c| in \verb|java.util|.

For example, \texttt{Data.Set}~\cite{hackage:containers-set} does this, so we can
make a set \texttt{s\ =\ S.fromList\ {[}False,\ True{]}} and then write
\texttt{and\ s}. It will produce the same result as the expression
\verb|and (toList s)|.

Why all this? Actually, this is not about \texttt{toList}. Quite the
contrary - this \texttt{Foldable} class exists to make it easier to
\emph{avoid} lists. Typically, we would not implement or call the
\texttt{toList} function, but use \texttt{foldMap} or \texttt{foldr}
instead --- or functions (like \verb|and|) that are
defined via \verb|foldMap|.

We give another example. Ancient libraries used to have a
function with name and type
\verb|sum :: Num a => [a] -> a|
but now its actual type is different:
\verb|sum :: (Num a, Foldable t) => t a -> a|,
and so we can take the sum of all elements in a set
\verb|s :: Set Int| simply via \verb|sum s|, instead of writing
\verb|sum (toList s)|. See - the lists are gone from the source
code.

Let us look at the type of
\href{https://hackage.haskell.org/package/base-4.9.1.0/docs/Data-Foldable.html\#v:foldMap}{\texttt{foldMap}},
and the class
\href{https://hackage.haskell.org/package/base-4.9.1.0/docs/Data-Monoid.html\#t:Monoid}{\texttt{Monoid}}
that it uses.

\begin{verbatim}
class Foldable t where
    foldMap :: Monoid m => (a -> m) -> t a -> m

class Monoid m where
    mempty :: m ; mappend :: m -> m -> m
\end{verbatim}

This is fancy syntax for the following (slightly re-arranged)

\begin{verbatim}
foldMap :: m -> (a -> m) -> (m -> m -> m) -> t a -> m
\end{verbatim}
where the first and third argument are  the
dictionary for \verb|instance Monoid m|.

So when you call \texttt{foldMap} over some collection (of type
\texttt{t\ a}), then you have to supply (via the \verb|Monoid| instance)
the result value
(\texttt{mempty\ ::\ m}) for the empty collection, the mapping (of type
\verb|a -> m|) to be applied to individual elements,
and a binary operation
\verb|mappend :: m -> m -> m| to be
used to combine results for sub-collections.

Let us see how this is used to \texttt{sum} over \texttt{Data.Set}, the
type mentioned above, internally defined as size-balanced search trees in
\verb|Data.Set.Internal|~\cite{hackage:containers-set}
and repeated here for convenience:

\begin{verbatim}
data Set a = Bin Size a (Set a) (Set a) | Tip

instance Foldable Set where
  foldMap f t = go t
    where go Tip = mempty
          go (Bin 1 k _ _) = f k
          go (Bin _ k l r) = go l `mappend` (f k `mappend` go r)
\end{verbatim}

This means that \texttt{foldMap\ f\ s}, for
\texttt{s\ ::\ Set\ a} of size larger than one, does the following: each
branch node \texttt{n} is replaced by two calls of \texttt{mappend} that
combine results of recursive calls in subtrees \texttt{l} and \texttt{r}
of \texttt{n}, and \texttt{f\ k} for the key \texttt{k} of \texttt{n},
and we make sure to traverse the left subtree, then the key at the root,
then the right subtree.

The definition~\cite{hackage:foldable} of \verb|sum| is

\begin{verbatim}
class Foldable t where
  sum :: Num a => t a -> a
  sum = getSum #. foldMap Sum
\end{verbatim}

with the following auxiliary types~\cite{hackage:monoid}

\begin{verbatim}
newtype Sum a = Sum { getSum :: a }
instance Num a => Monoid (Sum a) where
    mempty = Sum 0 ; mappend (Sum x) (Sum y) = Sum (x + y)
\end{verbatim}

So, when we take the \texttt{sum} of some \texttt{Set\ a}, we do
a computation that adds up all elements of the set, but without an
explicit iterator - there is no list in sight.

\section{Discussion and Related Opinions}\label{discuss-this}

\paragraph{Specifically for Section~\ref{lists-that-are-really-not-there}}:
When the types for \texttt{sum,\ and}, and many more functions
were generalized from lists
to \texttt{Foldable} a while back, some \texttt{Foldable}
instances were introduced that result in funny-looking semantics like
\texttt{maximum\ (2,1)\ ==\ 1}, which is only ``natural'' if you think
of the curried type constructor application \texttt{(,)\ a\ b},
remove the last argument, and demand that \texttt{(,)\ a} ``obviously''
must be \texttt{Functor}, \texttt{Foldable}, and so on.
This might save some keystrokes, but is has been
criticized~\cite{thielemann:haskell-matlab}
because it ``moves Haskell into the Matlab league of languages
where any code is accepted, with unpredictable results.''

\paragraph{In general}: This text is based on my experience
in teaching (advanced) programming language concepts, for two decades now,
and using Haskell for real-world code, e.g.,
an E-Learning/Assessment system~\cite{waldmann:autotool-fdpe02,autotool}
in use since 2001,
and an administration and presentation layer
for Termination Competitions~\cite{star-exec-presenter}.

Other researchers and practitioners have expressed similar opinion
on using (or, avoiding) lists.
Breitner teaches a course~\cite{breitner:cis194} that is similar to mine,
and Thielemann recommends~\cite{thielemann:lists-are-not-good}
to ``choose types properly'', in particular,
to not confuse lists with arrays.

\paragraph{Previous discussion of this text}: see
\url{https://mail.haskell.org/pipermail/haskell-cafe/2017-March/thread.html\#126457}
and
\url{https://www.reddit.com/r/haskell/comments/5yiusn/when_you_should_use_lists_in_haskell_mostly_you/}.

\bibliographystyle{alpha}
\bibliography{list-or-not-list}

\begin{thebibliography}{WMvdK14}

\bibitem[Aug97]{augustsson:heap-sort}
Lennart Augustsson.
\newblock {Re: heap sort or the wonder of abstraction}.
\newblock messsage on mailing list
  \url{https://www.mail-archive.com/haskell@haskell.org/msg01822.html}, 1997.

\bibitem[Bre16]{breitner:cis194}
Joachim Breitner.
\newblock {CIS 194: Introduction to Haskell (Fall 2016)}.
\newblock \url{http://cis.upenn.edu/~cis194/fall16/}, 2016.

\bibitem[GHJV95]{gof:design-patterns}
Erich Gamma, Richard Helm, Ralph Johnson, and John Vlissides.
\newblock {\em {Design Patterns: Elements of Reusable Object-Oriented
  Software}}.
\newblock Addison-Wesley, 1995.

\bibitem[Gib06]{gibbard:lists-are-loops}
Cale Gibbard.
\newblock {Export lists in modules}.
\newblock message on mailing list
  \url{https://mail.haskell.org/pipermail/haskell-prime/2006-February/000795.html},
  2006.

\bibitem[Gil01]{hackage:monoid}
Andy Gill.
\newblock {\texttt{Data.Monoid}: A class for monoids with various
  general-purpose instances}.
\newblock
  \url{https://hackage.haskell.org/package/base-4.11.1.0/docs/Data-Monoid.html},
  2001.

\bibitem[GLP93]{DBLP:conf/fpca/GillLJ93}
Andrew~John Gill, John Launchbury, and Simon~L. {Peyton Jones}.
\newblock {A Short Cut to Deforestation}.
\newblock In John Williams, editor, {\em Proc. Conf. on Functional programming
  languages and computer architecture, {FPCA} 1993}, pages 223--232. {ACM},
  1993.

\bibitem[Gra01]{graham:roots-of-lisp}
Paul Graham.
\newblock {The Roots of LISP}.
\newblock \url{http://www.paulgraham.com/rootsoflisp.html}, 2001.

\bibitem[Hin69]{hindley:principal}
J.~Roger Hindley.
\newblock {The Principal Type-Scheme of an Object in Combinatory Logic}.
\newblock {\em Transactions of the American Mathematical Society}, 146:29--60,
  1969.

\bibitem[Lei02]{hackage:containers-set}
Daan Leijen.
\newblock {\texttt{Data.Set}: Finite Sets, part of the containers library}.
\newblock
  \url{https://hackage.haskell.org/package/containers-0.6.0.1/docs/Data-Set.html},
  2002.

\bibitem[Les10]{hackage:vector}
Roman Leshchinskiy.
\newblock {\texttt{Data.Vector}: Efficient Arrays}.
\newblock
  \url{https://hackage.haskell.org/package/vector-0.12.0.1/docs/Data-Vector.html},
  2010.

\bibitem[Mar10]{haskellreport}
Simon Marlow.
\newblock {Haskell 2010 Language Report}.
\newblock \url{https://www.haskell.org/onlinereport/haskell2010/}, 2010.

\bibitem[McC60]{DBLP:journals/cacm/McCarthy60}
John McCarthy.
\newblock {Recursive Functions of Symbolic Expressions and Their Computation by
  Machine, Part {I}}.
\newblock {\em Commun. {ACM}}, 3(4):184--195, 1960.

\bibitem[Nor07]{norell:thesis}
Ulf Norell.
\newblock {\em Towards a practical programming language based on dependent type
  theory}.
\newblock PhD thesis, Department of Computer Science and Engineering, Chalmers
  University of Technology, SE-412 96 G\"{o}teborg, Sweden, September 2007.

\bibitem[{Ora}18]{java.util.List}
{Oracle}.
\newblock {Interface \texttt{List<E>}, Java Platform Version 10 API
  Specification}.
\newblock \url{https://docs.oracle.com/javase/10/docs/api/java/util/List.html},
  2018.

\bibitem[Pat05]{hackage:foldable}
Ross Paterson.
\newblock {\texttt{Data.Foldable}: class of data structures that can be folded
  to a summary value}.
\newblock
  \url{https://hackage.haskell.org/package/base-4.11.1.0/docs/Data-Foldable.html},
  2005.

\bibitem[RW02]{waldmann:autotool-fdpe02}
Mirko Rahn and Johannes Waldmann.
\newblock {The Leipzig autotool System for Grading Student Homework}.
\newblock In {\em Workshop Functional and Declarative Programming in Education
  (FDPE 2002)}, 2002.

\bibitem[Thi07]{thielemann:lists-are-not-good}
Henning Thielemann.
\newblock Choose types properly---lists are not good for everything.
\newblock
  \url{https://wiki.haskell.org/Haskell_programming_tips\#Lists_are_not_good_for_everything},
  2007.

\bibitem[Thi16]{thielemann:haskell-matlab}
Henning Thielemann.
\newblock {Haskell Foldable Wats}.
\newblock message on mailing list
  \url{https://mail.haskell.org/pipermail/libraries/2016-February/026678.html},
  2016.

\bibitem[Tur87]{DBLP:journals/eatcs/Turner87}
David Turner.
\newblock An overview of miranda.
\newblock {\em Bulletin of the {EATCS}}, 33:103--114, 1987.

\bibitem[Tur12]{DBLP:conf/sfp/Turner12}
D.~A. Turner.
\newblock {Some History of Functional Programming Languages - (Invited Talk)}.
\newblock In Hans{-}Wolfgang Loidl and Ricardo Pe{\~{n}}a, editors, {\em Trends
  in Functional Programming (TFP)}, volume 7829 of {\em Lecture Notes in
  Computer Science}, pages 1--20. Springer, 2012.

\bibitem[{Uni}01]{hackage:list}
{University of Glasgow}.
\newblock {Data.List: Operations on Lists}.
\newblock
  \url{https://hackage.haskell.org/package/base-4.11.1.0/docs/Data-List.html#g:21},
  2001.

\bibitem[Wad90]{DBLP:journals/tcs/Wadler90}
Philip Wadler.
\newblock {Deforestation: Transforming Programs to Eliminate Trees}.
\newblock {\em Theor. Comput. Sci.}, 73(2):231--248, 1990.

\bibitem[Wal02]{autotool}
Johannes Waldmann.
\newblock Leipzig autotool.
\newblock
  \url{https://gitlab.imn.htwk-leipzig.de/autotool/all0#leipzig-autotool},
  2002.

\bibitem[Wal17a]{waldmann:data}
Johannes Waldmann.
\newblock {How Do We Store Data, Then?}
\newblock \url{https://www.imn.htwk-leipzig.de/~waldmann/etc/untutorial/data/},
  2017.

\bibitem[Wal17b]{waldmann:how-i-teach}
Johannes Waldmann.
\newblock {How I Teach Functional Programming}.
\newblock In {\em Workshop on Functional Logic Programming (WFLP 2017),
  W\"urzburg}, 2017.

\bibitem[WMvdK14]{star-exec-presenter}
Johannes Waldmann, Rene Muhl, and Stefan von~der Krone.
\newblock {star-exec-presenter: A Web Application for Displaying Results of the
  Termination Competition}.
\newblock \url{https://github.com/jwaldmann/star-exec-presenter}, 2014.

\end{thebibliography}

\end{document}